# Investigation Into the Viability of Neural Networks as a Means for Anomaly Detection in Experiments Like Atlas at the LHC

**Sully Billingsley**

BS in Physics at the University of Texas at Arlington/ Arlington TX, US

<cole.billingsley@mavs.uta.edu>

## Abstract

Petabytes of data are generated at the Atlas experiment at the Large Hadron Collider however not all of it is necessarily interesting, so what do we do with all of this data and how do we find these interesting needles in an uninteresting haystack. This problem can possibly be solved through the process of anomaly detection. In this document, Investigation Into the Viability of Neural Networks as a Means for Anomaly Detection in Experiments Like Atlas at the LHC the effectiveness of different types of neural network architectures as anomaly detectors are researched using Monte Carlo simulated data generated by the DarkMachines project. This data is meant to replicate Standard Model and Beyond Standard Model events. By finding an effective model, the Atlas experiment can become more effective and fewer interesting events will be lost.

## 1  Problem Statement

Anomaly detection is important many contexts, it is utilized to identify events or items in sets that show large contrast to the normal. For example, detecting malicious web traffic, identifying heart arrhythmia, fraudulent purchases, and of course identifying Beyond Standard Model events in the Atlas experiment. Not all anomaly detection utilizes neural networks, however this study will. Some other means for anomaly detection are K-means, random forest, isolation forest, and dimensionality reduction. Neural networks will be used in this study because they have shown promise in other studies, [2], [4], and [5], and their generality in other uses. The modularity of neural networks, specifically the Keras library, also helps in the choice because it will make it very simple to change model architecture. The data used gives the title of processes and descriptions of the particles that make them up. To identify anomalies the neural networks will be identifying processes that either fit the Standard Model or not.

## 2 Data Description

Figure 1. Standard Model processes:
$$pp \to jj$$
$$pp \to W^{\pm}(+2j)$$
$$pp \to \gamma(+2j)$$
$$pp \to Z(+2j)$$
$$pp \to t\bar{t}(+2j)$$
$$pp \to W^{\pm}t(+2j)$$
$$pp \to W^{\pm}\bar{t}(+2j)$$
$$pp \to W^{+}W^{-}(+2j)$$
$$pp \to t+\text{jets}(+2j)$$
$$pp \to \bar{t}+\text{jets}(+2j)$$
$$pp \to \gamma\gamma(+2j)$$
$$pp \to W^{\pm}\gamma(+2j)$$
$$pp \to ZW^{\pm}(+2j)$$
$$pp \to Z\gamma(+2j)$$
$$pp \to ZZ(+2j)$$
$$pp \to h(+2j)$$
$$pp \to t\bar{t}\gamma(+1j)$$
$$pp \to t\bar{t}Z$$
$$pp \to t\bar{t}h(+1j)$$
$$pp \to \gamma t(+2j)$$
$$pp \to t\bar{t}W^{\pm}$$
$$pp \to \gamma\bar{t}(+2j)$$
$$pp \to Zt(+2j)$$
$$pp \to Z\bar{t}(+2j)$$
$$pp \to t\bar{t}t\bar{t}$$
$$pp \to t\bar{t}W^{+}W^{-}$$

*Figure 1. Standard Model processes*

To prove the viability of the neural networks, they will be tested on data simulated to replicate events at the Atlas Experiment at the Large Hadron Collider. In Figure 1, the possible events for the Standard Model are listed. In Figure 2, the possible events for the Beyond Standard Model are described. The data is in the format of a CSV and each row a separate event. The row describes the transverse energy, the angle of the center of mass of the collision, and the 4-vector of all the particles that make up the event. From this list of events we can train a neural network to reconstruct these events and classify them as either Standard Model or Beyond Standard Model.

To visualize the data histograms have been generated, summarizing certain elements of the data. In the histograms the transverse momenta, transverse energy, pseudorapidity (angle of the particle relative to the beam axis), and azimuthal scattering angle are depicted for both the jets and leptons in the events. The histograms are color-coded based on the process. For example, proton-proton to jet-jet can be seen in a dark red and is labeled njets, which is the name of the file those events with that process were saved under. The last two histograms show the missing transverse energy and the missing transverse angle for each event in the background processes. Pseudorapidity is denoted with the Greek symbol eta, η, and azimuthal angle is denoted with the Greek symbol phi, ϕ.

Figure 2. Beyond Standard Model processes:
$$pp \to \tilde{g}\tilde{g}\,(1\,\text{TeV})$$
$$pp \to \tilde{g}\tilde{g}\,(1.2\,\text{TeV})$$
$$pp \to \tilde{g}\tilde{g}\,(1.4\,\text{TeV})$$
$$pp \to \tilde{g}\tilde{g}\,(1.6\,\text{TeV})$$
$$pp \to \tilde{g}\tilde{g}\,(1.8\,\text{TeV})$$
$$pp \to \tilde{g}\tilde{g}\,(2\,\text{TeV})$$
$$pp \to \tilde{g}\tilde{g}\,(2.2\,\text{TeV})$$
$$pp \to \tilde{t}_1\bar{\tilde{t}}_1\,(220\,\text{GeV}), m_{\tilde{\chi}_1^0} = 20\,\text{GeV}$$
$$pp \to \tilde{t}_1\bar{\tilde{t}}_1\,(300\,\text{GeV}), m_{\tilde{\chi}_1^0} = 100\,\text{GeV}$$
$$pp \to \tilde{t}_1\bar{\tilde{t}}_1\,(400\,\text{GeV}), m_{\tilde{\chi}_1^0} = 100\,\text{GeV}$$
$$pp \to \tilde{t}_1\bar{\tilde{t}}_1\,(800\,\text{GeV}), m_{\tilde{\chi}_1^0} = 100\,\text{GeV}$$
$$pp \to Z'\,(2\,\text{TeV})$$
$$pp \to Z'\,(2.5\,\text{TeV})$$
$$pp \to Z'\,(3\,\text{TeV})$$
$$pp \to Z'\,(3.5\,\text{TeV})$$
$$pp \to Z'\,(4\,\text{TeV})$$

*Figure 2. Beyond Standard Model processes*

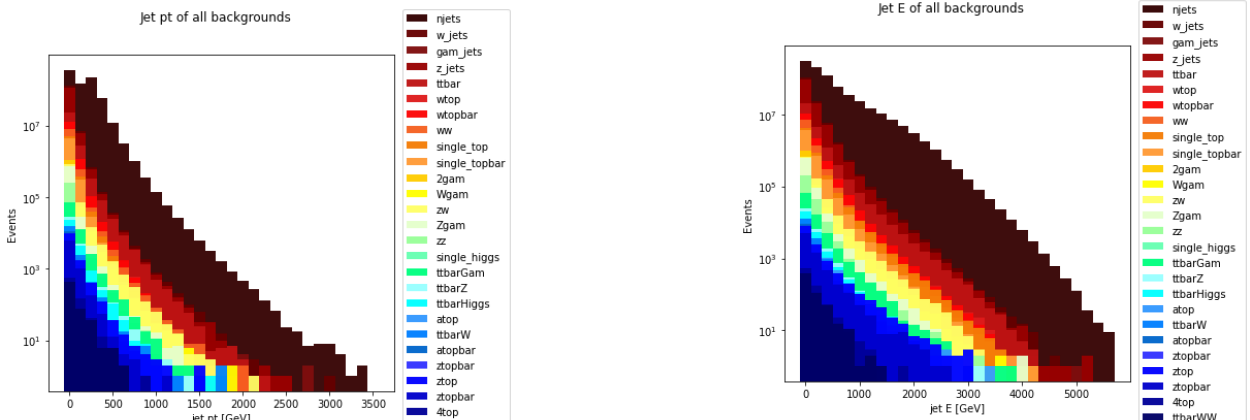

*Figure 3. Transverse momenta and energy of all the jets in background processes*

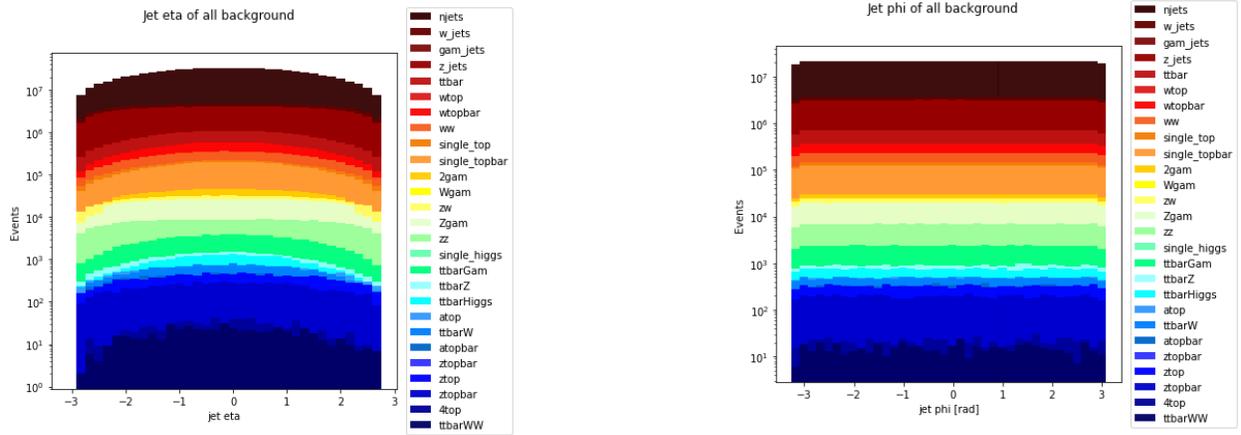

*Figure 4. Pseudorapidity and azimuthal angle of jets in background processes*

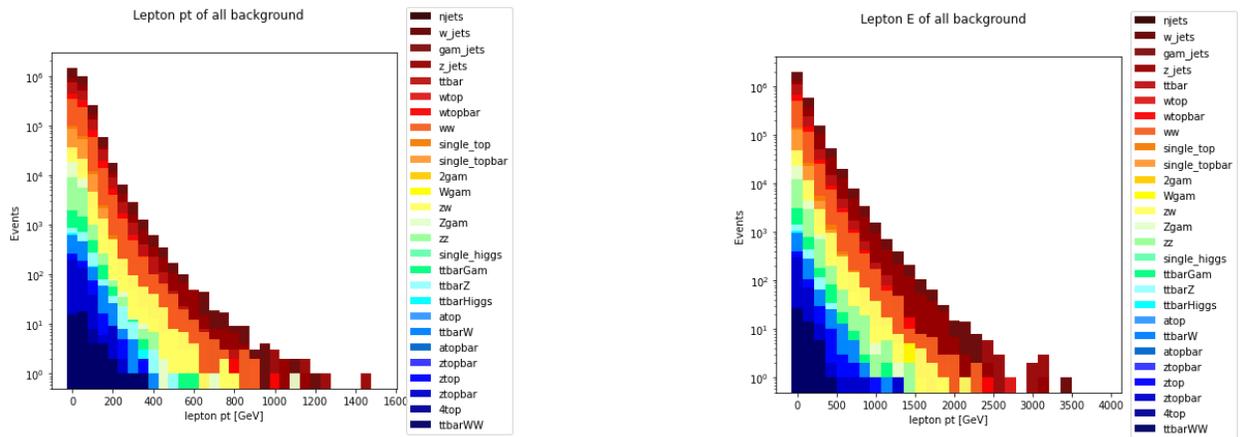

*Figure 5. Transverse momenta and energy of all the leptons in background processes*

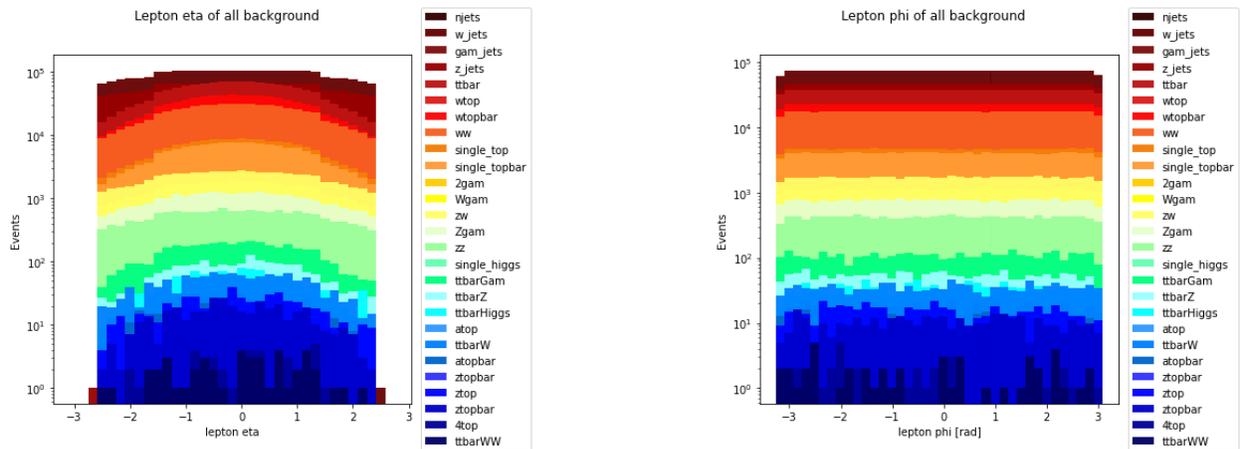

*Figure 6. Pseudorapidity and azimuthal angle of all the leptons in background processes*

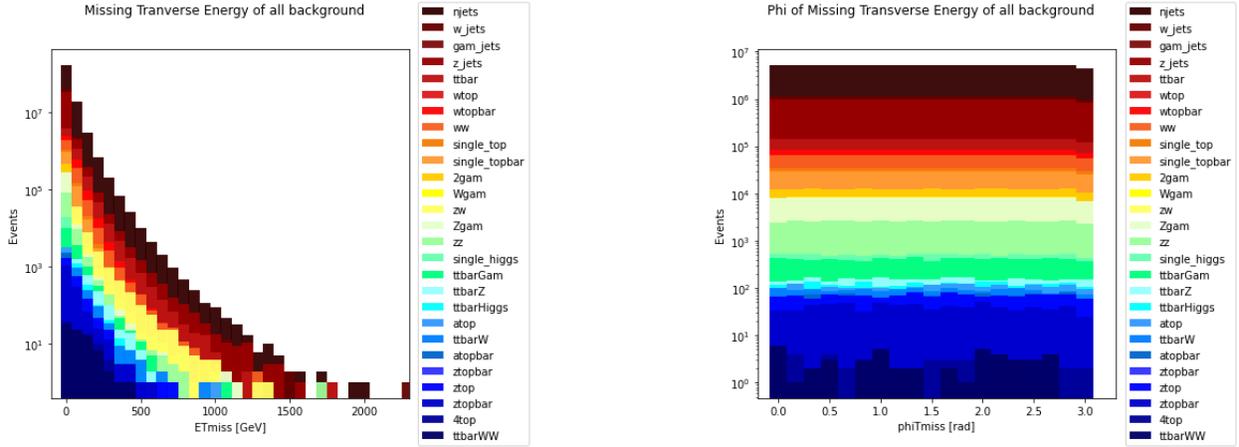

*Figure 7. Missing transverse energy and azimuthal angle for all background*

# 3 Approach

For the reader to better understand the neural networks and how they can be effective for anomaly detection and classification they can turn their attention to [4]. However, for a brief introduction, neural networks can be understood as a mathematical model that emulates the connections made in a brain. The weights of the neurons and connections change according to the loss between the current output and the corresponding input. As the loss gets lower the model's weights change in that direction.

For this specific application, multiple models will be made to increase the likelihood of finding a model that can effectively identify anomalies. The base of all these models will be a variational autoencoder (VAE), [2] and [4], which have been used in anomaly detection in the past and have shown to be effective. The VAE works by trying to reconstruct the input that is given. If the difference, or loss, is much higher than what the training loss was, this event will be flagged as an anomaly. A drawback from this approach is that information can be lost when going from the latent space, the most compressed information state, to the output. This area is known as the decoding section.

As a way to counteract this, a binary classifier can be made by attaching a single neuron with a sigmoid activation function to the latent space of a model. The anomaly will be identified through a Boolean logic statement. The two possibilities are not anomalous or anomalous. If the output from this single neuron is greater than .5, it is not anomalous and below is anomalous.

Now that the model types have been defined, models can be created through training. The first model will be created by training a VAE on data from the Standard Model processes and Beyond Standard Model processes. There is an equal amount of each type to reduce bias. The second model is a binary classifier that is attached to the latent space of the first model once the first model is already trained. The third model is another VAE that is trained on only Standard Model processes. The last is a binary classifier similar to the second model but is attached the latent space of the third model once the third model is trained.

# 4      Results

Both VAE's were trained in the same fashion apart from the data they were trained on. They were trained for 100 epochs on a batch size of 256 samples and the best model was saved depending on the validation loss. The binary classifiers were given one epoch of training on mixed data to remove the random nature of the initialization of neurons in Keras. The section that the binary classifier was attached to was not allowed to have its weights change during this training to preserve the information of the training of the VAE's. The accuracy of all the models was determined on the same set of validation data that none of the models had been trained on.

Figure 8 shows the loss during training of the first model. It is on a logarithmic chart to show greater detail of the loss. It can be seen that the loss curve had started to level off but was not completely level so more training could have been completed but limitations in computation time prevented this. More training epochs could be used in further training.

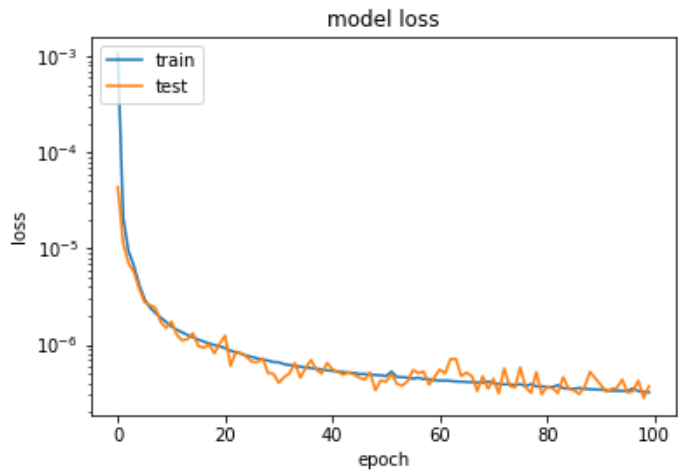

*Figure 8. Training loss of VAE train on mixed data*

Figure 9 gives a visualization of how well the model can reconstruct the background and signal. The greater the area under the curve of the histogram, the worse the model can reconstruct the given input. It can be seen that the model performs poorly when trying to identify anomalies through reconstruction loss. This makes sense because the model was trained to reconstruct both background and signal. Its performance can be visualized in the plot on the right in Figure 9. The area under the curve was .7065 which translates to a 70.65% accuracy in identifying anomalies.

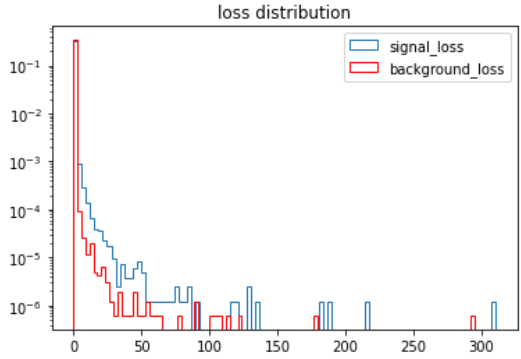 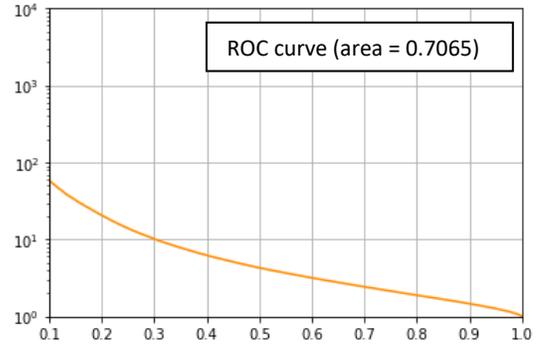

*Figure 9. Loss distribution and accuracy curve of VAE trained on mixed data*

Figure 10 shows the false positive rate vs. the true positive rate and the area under this curve gives the accuracy of the second model. In this case, the accuracy was 99.362%.

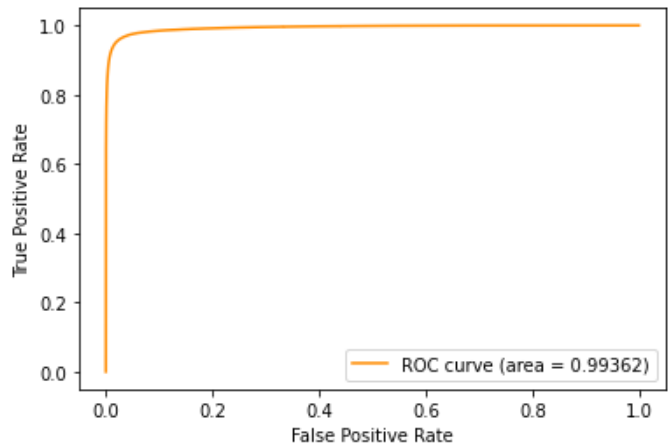

*Figure 10. Accuracy curve of binary classifier trained on mixed data*

Figure 11 shows the same information Figure 8 does but pertaining to the third model, the VAE trained on background events only.

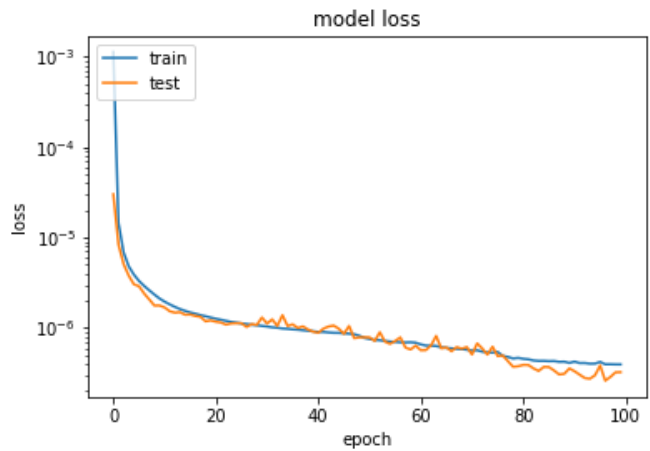

*Figure 11. Training loss for VAE on background events*

Figure 12 reveals interesting results. It can be seen that the third model can not do a good job reconstructing the signal inputs. This is derived by the significant difference in the area under the curve between the two histograms in the loss distribution chart. This large discrepancy between the two makes the accuracy for this VAE much higher than the first model. This is because the losses of the signal are on average much higher than the background making it easy for a selection for an anomaly easy.

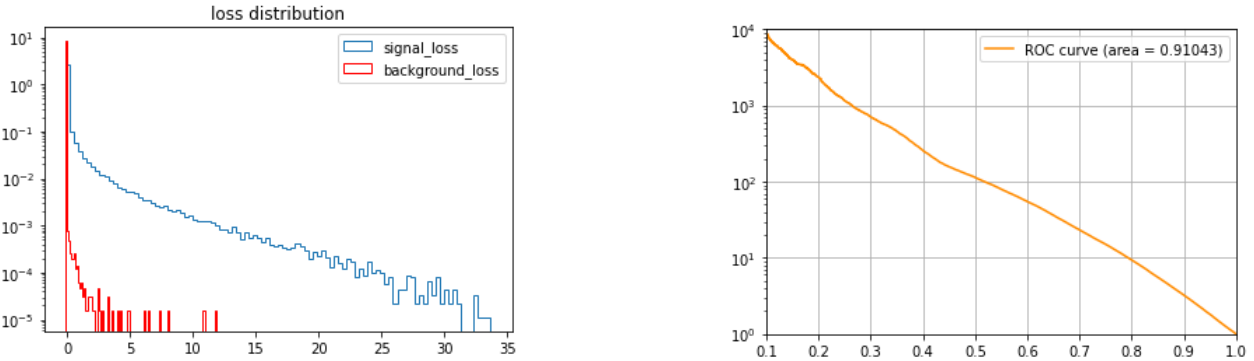
*Figure 12. Loss distribution and accuracy curve for VAE trained on background events*

Figure 13 is analogous to Figure 10 but pertains to the fourth model and had an accuracy 99.329%.

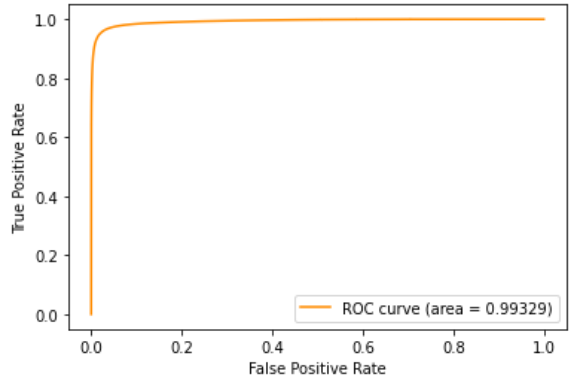
*Figure 13. Accuracy curve of binary classifier trained on background data*

## 5    Conclusion

As shown in the model performance, using a neural network as a means to identify anomalies is an extremely viable option. Three out of four of the models performed with greater than 90% accuracy. The binary classifiers in both cases performed the best. This is because when the model is trying to reconstruct the input, it loses information that help it make its decision. Therefore the best choice to detect anomalies in this case would be the binary classifier that was trained on background events because this model does not need to define the events that are beyond the Standard Model making it accurate in more cases. This means that if this model were implemented at the Large Hadron Collider fewer interesting events will be lost, further increasing the efficacy of the experiment

## Acknowledgements

This project was completed with the guidance of Dr. Amir Farbin and the data generated from the DarkMachines project. This work is also a continuation of that of Dr. Debottam Bakshigupta and Sergio Garza.